\documentclass[manuscript]{aastex}

\def\Msun{$\textit{M}_{\odot}$}
\def\Lsun{$\textit{L}_{\odot}$}
\def\Rsun{$\textit{R}_{\odot}$}

\newcommand{\kms}{km s$^{-1}$}
\newcommand{\ha}{H$\alpha$} 
\newcommand{\hb}{H$\beta$} 
\newcommand{\lam}{$\lambda$}

\shorttitle{Spectra of HBC\,722}
\shortauthors{Lee et al.}

\begin{document}

\title{High Resolution Optical and NIR Spectra of HBC\,722
\footnote{Based on observations obtained with the Hobby-Eberly Telescope,
which is a joint project of the University of Texas at Austin, the
Pennsylvania State University, Stanford University,
Ludwig-Maximilians-Universit\"{a}t M\"{u}nchen, and
Georg-August-Universit\"{a}t G\"{o}ttingen.}
}

\author{Jeong-Eun Lee$^{1,2}$, Sunkyung Park$^1$, Joel D. Green$^{2,3}$, William D. Cochran$^2$, Wonseok Kang$^4$, Sang-Gak Lee$^4$, and Hyun-Il Sung$^5$} 
\affil{
	$^1$ School of Space Research, Kyung Hee University, 1 Seocheon-dong, Giheung-gu, Yongin-si, Gyeonggi-do 446-701, Korea; 
	    jeongeun.lee@khu.ac.kr, sunkyung@khu.ac.kr  \\
	$^2$ Department of Astronomy, University of Texas at Austin, TX, USA,;
	    joel@astro.as.utexas.edu, wdc@astro.as.utexas.edu \\
	 $^3$ Space Telescope Science Institute, Baltimore, MD, USA,\\    
	$^4$ National Youth Space Center, 11-1, Deokheung-ri, Dongil-myeon, Goheung-gu, Jeollanam-do 548-951, Republic of Korea;
	    wskang@nysc.or.kr, sanggak@kywa.or.kr \\	    
\	$^5$ Korea Astronomy and Space Science Institute, 36-1 Whaam-dong, Yuseong-gu, Daejon 305-348, Republic of Korea; 
	    hisung@kasi.re.kr \\
	    }

\begin{abstract}

We present the results of high resolution (R$\ge$30,000) optical and near-IR spectroscopic monitoring observations of HBC\,722, a recent FU Orionis object that underwent an accretion burst
in 2010.  We observed HBC\,722 in optical/near-IR with the BOES, HET-HRS, and IGRINS spectrographs, at various points in the outburst.  We found atomic lines with strongly blueshifted absorption features or P Cygni profiles, both evidence of a wind driven by the accretion. Some lines show a broad double-peaked absorption feature, evidence of disk rotation. However, the wind-driven and disk-driven spectroscopic features are anti-correlated in time; the disk features became strong as the wind features disappeared. This anti-correlation might indicate that the rebuilding of the inner disk was interrupted by the wind pressure during the first two years.  The Half-Width at Half-Depth (HWHD) of the double-peaked profiles decreases with wavelength, indicative of the Keplerian rotation; the optical spectra with the disk feature are fitted by a G5 template stellar spectrum convolved with a rotation velocity of 70 \kms\ while the near-IR disk features are fitted by a K5 template stellar spectrum convolved with a rotation velocity of 50 \kms.  Therefore, the optical and near-IR spectra seem to trace the disk at 39 and 76 \Rsun, respectively.  We fit a power-law temperature distribution in the disk, finding an index of 0.8, comparable to optically thick accretion disk models. 

\end{abstract}


\section{Introduction}
During star formation, accretion onto protostars through disks seems to occur episodically (Audard et al. 2014 and references therein), and FU Orionis-type objects (hereafter, FUors) have been proposed as prominent examples of burst accreting protostars.
HBC\,722,  a new member of the FUor group, brightened by about 5 magnitudes (in V) in the latter half of 2010 \citep{semkov10}. HBC\,722 reached a relative maximum brightness in September 2010 and  then faded until May 2011 \citep[e.g.][]{sung13}.  Late in 2011 the brightness began to increase gradually and reached its second maximum in the first half of 2013, after which it remained at approximately constant level. The object shows short-term brightness variations with several day periods \citep{green13,baek15}. The broad observational coverage and clear delineation of several epochs within the optical light curve make 
HBC\,722 a good case study of the accretion process and its effect on low mass star formation. 

Followup observations were not restricted to optical variability.  \citet{miller11} characterized the source with TripleSpec in the near-IR, and observed a typical spectrum for a T Tauri star, with a high accretion rate.  \citet{green11} observed the source with Herschel spectroscopy (at submillimeter wavelengths) and \citet{dunham12} observed the source with the SMA at millimeter wavelengths; both concluded that the source was a relatively evolved young star, lacking a significant extended envelope, and a modest accretion rate (few $\times$ 10$^{-6}$ \Msun yr$^{-1}$) by FUor standards \citep{kospal11}.  \citet{liebhart14} observed HBC\,722 with Chandra, and found that the X-ray increased during the burst, concluding that dust-free gas had been ejected by the star sometime after the early outburst.  Taken together, this suggests that the source was a low mass Class II (T Tauri) star pre-outburst, and it underwent a sudden accretion event that initiated a jet sometime around or after the initial optical burst.

The triggering event for FUors is not fully understood; in particular it is unclear whether the burst proceeds from the very innermost regions of the circumstellar disk, or whether it begins as a perturbation at larger disk radii following the formation and release of a disk instability \citep{zhu07}.  Optical/near-IR variability can trace inner disk activity but is difficult to distinguish from the activity on or near the stellar surface (e.g. flickering; e.g., \citealt{kenyon00}).  It is only with high resolution spectroscopy, covering the earliest stages of burst, that we can distinguish variation in the disk from variations near the star and outflow/jet.

In this paper, we report monitoring spectroscopic observations of HBC\,722 with high resolution spectrometers in optical and near-IR.  We use higher cadence but lower signal-to-noise spectroscopy to guide our less frequent high signal-to-noise spectroscopic observations.  We find a highly blueshifted wind component in absorption in the profiles of various lines including  \ha\ and Ca II \lam 8498, which show a P Cygni profile.
We also observe a broad double-peaked absorption feature, which is an evidence of the disk rotation \citep{hartmann96, zhu09}, in the neutral metal lines both at the optical and near-IR wavelengths.  These double-peaked features imply a rotating Keplerian inner disk (within $<$ 1 AU) for which we can constrain the current temperature profile.

\section{Observations}
\subsection{HET - HRS}
We observed HBC\,722 with the High Resolution Spectrograph
\citep{tull98} of the Hobby-Eberly Telescope \citep{ramsey98}
at McDonlad Observatory on the nights of 07 December 2010,
31 May 2011, 03 August 2012 and 30 August 2012, 30 May 2013
and 08 August 2013.  These spectra were taken using the ``600g5822''
HRS configuration.   This provided spectral coverage from 479 to 681\,nm
at a resolving power of $R = \lambda/\Delta\lambda =  30,000$.
The target was imaged onto a 2\,arcsec fiber, and two additional
2\,arcsec fibers were used to gather light from the sky immediately
adjacent to the target.  The CCD detectors were bined $2 \times 2$
pixels to give four binned pixels per resolution element.
On each night, four separate spectra, each of 900\,second duration,
were obtained, for a total exposure time of 60\,minutes.
Data reductions were performed with the IRAF echelle package.
The four separate spectral images of HBC\,722 were added together,
using a `cosmic-ray' rejection algorithm in order to remove noise
spikes due to radiation hits on the CCDs.
Spectra of HBC\,722 and each of the two sky fibers were subtracted
separately.   The spectra from the two sky fibers were averaged,
and that spectrum was subtracted from the HBC\,722 spectrum. 
The scaling of the sky spectrum was determined empirically to
fully remove the Earth night-sky emission lines.

\subsection{BOAO - BOES}
The Bohyunsan Optical Echelle Spectrograph (BOES) is a high resolution echelle spectrograph 
attached to the 1.8-m optical telescope at Bohyunsan Optical Astronomy Observatory (BOAO).
HBC\,722 has been observed using BOES \citep{lee11} since November 2010 (see Table 1) 
with resolving power ($  R = \frac{\lambda} {\Delta\lambda} $) of 30,000 using the 300 $\mu$m fiber. 
The observed wavelength range was  3600\AA{} -- 10500\AA{}, covering the full optical spectrum. 
Two other FUors, FU Orionis and V1057 Cyg, have been observed with BOES on 26 January 2011 and 
11 September 2012, respectively. The spectra of standard stars used in Figure 1 were also observed 
with BOES.

The observational data were reduced by the IRAF (Image Reduction and Analysis Facility) {\tt echelle} package.
Each aperture from the spectral images was extracted using a master flatfield image. 
Using the flatfielding process, we corrected the interference fringes and pixel-to-pixel variations of the spectrum images.
A ThAr lamp spectrum was used for wavelength calibration.

\subsection{HJST - IGRINS}
HBC\,722 was observed with the Immersion Grating Infrared Spectrograph (IGRINS) installed at the 2.7 m Harlan J. Smith Telescope (HJST) at the McDonald observatory on 20 November, 2014.  IGRINS covers the full H and K bands with R=40,000 \citep{park14}. 
The spectra have been reduced by the IGRINS pipeline \footnote{https://github.com/igrins/plp}. 
A standard A0 star (HIP 103108) was observed immediately after the observation of HBC\,722 for the telluric line correction.

\section{Results}

\begin{figure}[!p]
\begin{center}
\epsscale{0.8}
\plotone{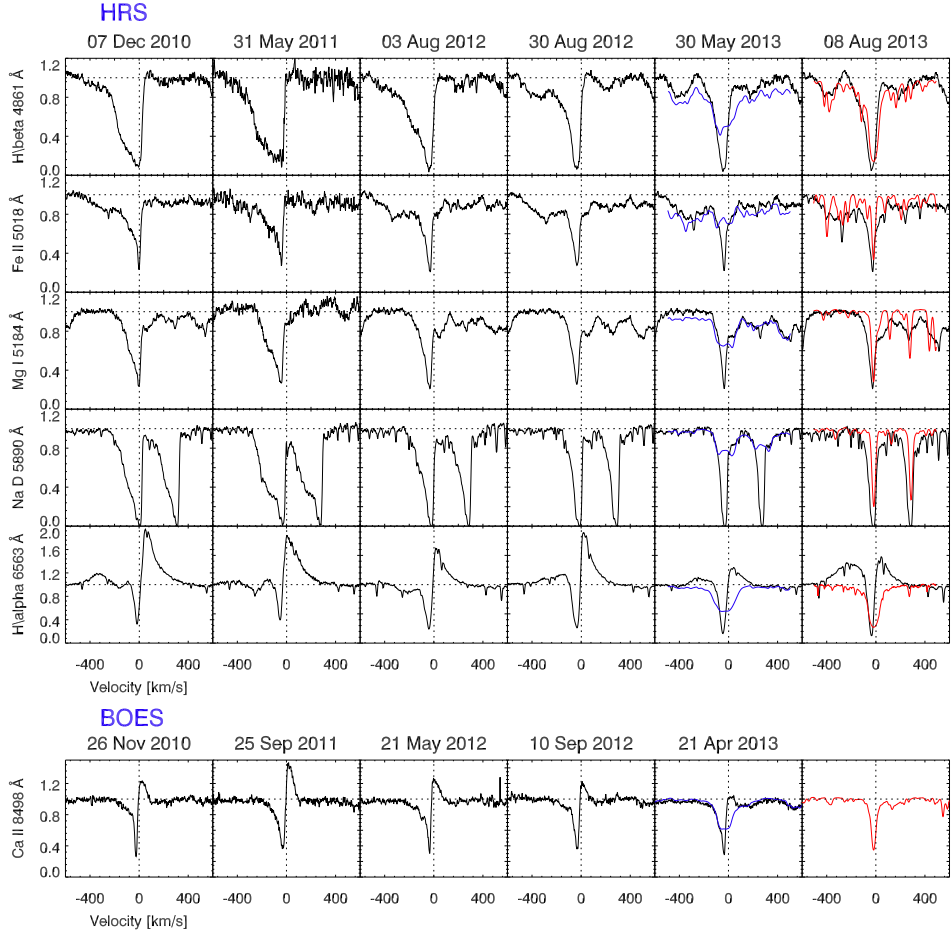}
\caption{\small The time variation of optical lines including the blueshifted (wind-tracing) absorption feature.  HET-HRS observations are presented in the upper panel, with five different lines shown as five rows, and six different observation epochs as six columns.  The BOES observations are presented in the lower panel in five observation epochs, which are close to the dates for HET-HRS observations.  
The red spectra in the boxes for August 2013 is the spectra of the G1 type star, HD 188650, which matches the best the deep narrow component of the spectra. The blue spectra in the boxes for May 2013 (April 2013 for BOES) is the spectra of the G5 type star, HD 52497, convolved with the rotational profile of $vsini=70$ \kms. The shallow broad component of each spectrum and the broadened adjacent features can be fitted well with these convolved spectra. 
} 
\label{timevar}
\end{center}
\end{figure}

In Figure \ref{timevar}, we compare the lines with a strong blue-shifted absorption feature with wings greater than $\sim$150 \kms, showing the profile variation throughout the burst.  Spectra observed from burst (2010) until 03 August 2012 show this feature, after which it disappears over the course of a single month. 

We observe a P Cygni (wind/outflow) profile in \ha\ and Ca II \lam 8498.  In \ha, we also observe a  redshifted emission wing as well as a blueshifted absorption at velocities as large as $\sim$400 \kms.  The Ca II profile is only covered by BOES  and does not appear as wide, but we cannot rule out a similar profile due to lower S/N in the BOES spectra. In Figure~\ref{timevar}, the HET-HRS spectra have much higher
signal-to-noise ratios compared to the BOES spectra so that we present only one BOES line, Ca II  at 8498 \AA, which is not covered by the HET-HRS.

If we examine the Na I D lines observed in May and August 2013 (Figure \ref{timevar}: 4th row, panels 5 and 6), the line is composed of a narrow deep component, typical of late F or G type stellar spectra, and a broad shallow  component, which we attribute to disk rotation. We compare this with a G1-type stellar spectrum (HD 188650; plotted in red) that matches the narrow deep component seen in the August 2013 spectra of HBC\,722. 

On first inspection, we also see a highly blueshifted shallow broad component  in \hb\ and Fe II 5018.  However, these features do {\it not} seem related to the blueshifted wind seen in \ha.  These additional blueshifted and redshifted absorption components appear in \hb\ and Fe II \lam 5018 (rows 1 and 2 of Figure \ref{timevar}).  Additional redshifted components only are noted in Mg I \lam 5184 (row 3 of Figure \ref{timevar}).

\begin{figure}[!p]
\begin{center}
\epsscale{0.8}
\plotone{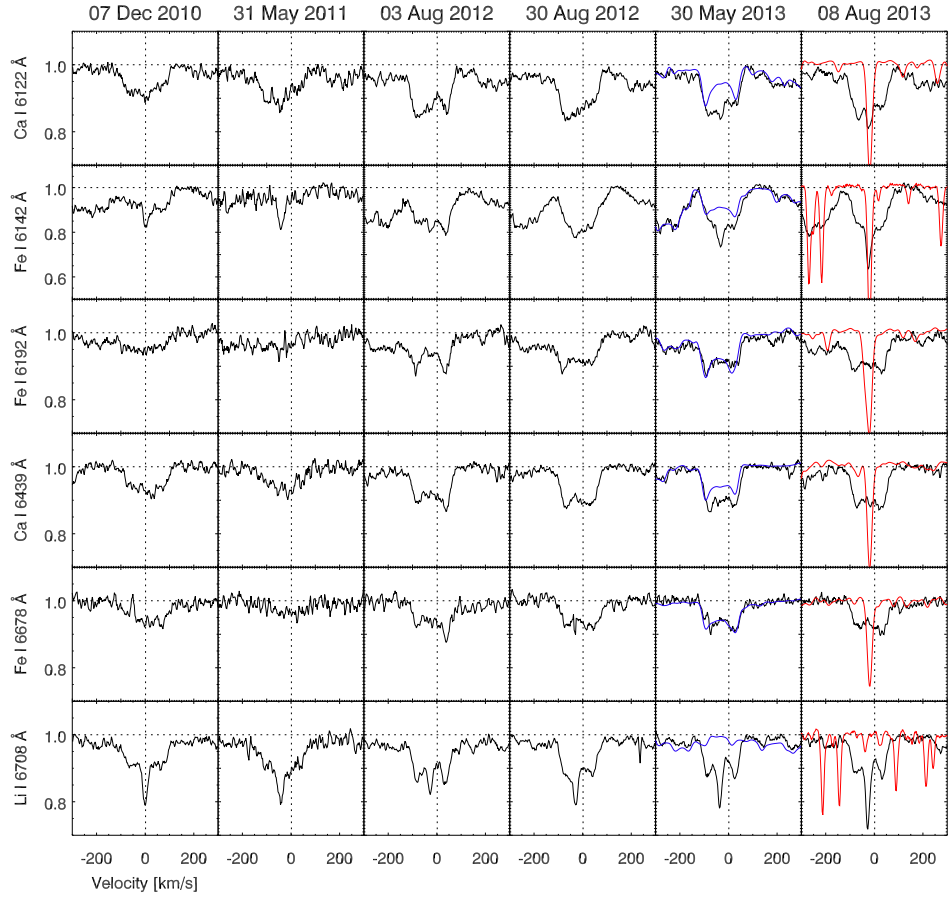}
\caption{Time variation of the double-peaked absorption features in various lines, evidence of disk rotation. Only the HET-HRS spectra are presented because the BOES spectra have much lower signal-to-noise ratio. The dates in the columns are the same as those in Figure~\ref{timevar}.
The blue and red spectra are also the same as those in Figure~\ref{timevar}.
} 
\label{doublepeak}
\end{center}
\end{figure}

These components show a broad flattened structure that is likely associated with other lines broadened by the rotation of the Keplerian disk.  These features are well matched with a G5 stellar template spectrum convolved with a rotational velocity ($vsini$) of 70 \kms\ as presented in blue together with the HBC\,722 spectra observed in May 2013.  This disk rotation is also evident in various neutral metal lines (Ca I, Fe I, Li I) as seen in Figure \ref{doublepeak}.  In these neutral lines, the high velocity blueshifted absorption seen in \ha\ is not present.

It has long been suggested that a double-peaked broad absorption feature in an FUor, like that seen in Figure \ref{doublepeak}, is clear evidence of disk rotation \citep{hartmann96, zhu09}.  Numerous metal lines in our spectra show this double-peaked profile; we list them along with measured HWHD \citep{petrov08} in Table 2. The average of HWHDs for optical lines is about 79 \kms, and their standard deviation is 7.6 \kms. It is not clear whether the scatter between individual lines inferred HWHDs may be significant.  To calculate HWHD, we must determine the  continuum level,  so the uncertainty of the continuum level as well as the different S/N ratios for individual lines can cause the observed scatter.

We fit these neutral metal line profiles by convolving a template G5 stellar spectrum with the rotational profile below. 
\begin{equation} \label{eq_rconvol}
\phi(\Delta{v})=\Big[1- \big(\frac{\Delta{v}} {v_{max}}\big)^{2}\Big]^{-1/2}, 
\end{equation}
where $v_{max} = vsini$ .
Most of the neutral metal lines are well-fitted by a G5-type stellar spectrum convolved with a rotational velocity of 70 \kms\ (shown in blue in the boxes for the spectra of HBC\,722 observed in May 2013, Figure \ref{doublepeak}).  

Many of these profiles gradually acquire a central narrow absorption feature after May 2013 (Figure \ref{doublepeak}); we attribute this feature to the central object, which has begun to contribute as the stellar brightness increased.  The stellar lines of Ca I \lam 6122, Fe I \lam 6142, and Li I \lam 6708 are very deep in August 2013.  The linewidth of the central feature is reasonably similar to the G1 template, but the disk features are similar to the G5 template convolved with a rotational profile.  One notable exception is Li \lam 6707 which is not observed in the G5 template but appears in HBC\,722.  The reverse is true for Ca I \lam 6439, which appears in the G5 template but not in HBC\,722.  

\begin{figure}[!p]
\begin{center}
\epsscale{0.8}
\plotone{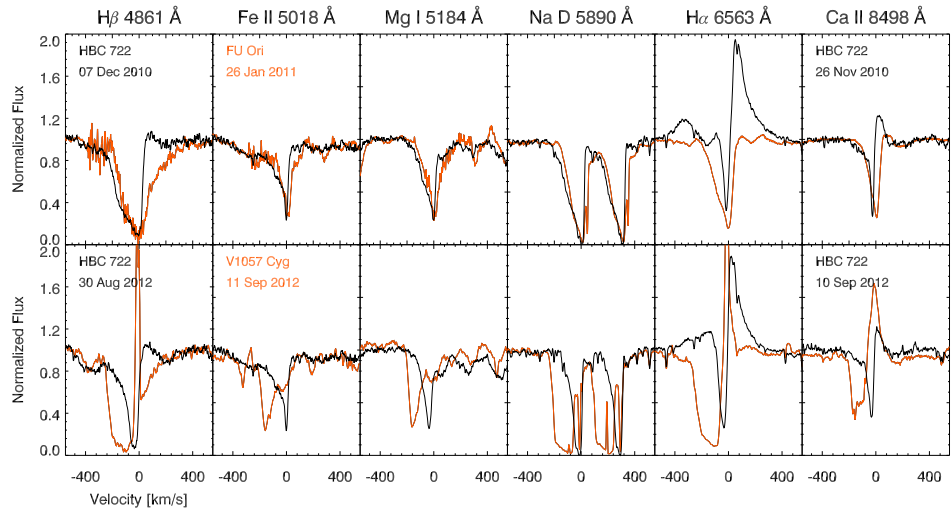}
\caption{Comparison of HBC\,722 in early outburst vs. FU Ori (top row), and later in outburst after the apparent wind-turnoff vs. V1057 Cyg (bottom row).  The HBC\,722 observations are from HET-HRS, with the exception of the Ca II line which was observed by BOES.  All observations of FU Ori and V1057 Cyg are from BOES. 
} 
\label{fuorscomp}
\end{center}
\end{figure}

We compare the spectra in early and late burst to FU Orionis and V1057 Cyg, two classical FUors, in Figure \ref{fuorscomp}.  Early in outburst, HBC\,722 looks very similar to FU Ori currently, with a strong blueshifted wind-driven feature and double-peaked absorption profiles.  FU Ori lacks a substantial circumstellar envelope \citep[e.g.][]{dunham12,green13b,audard14} as does HBC\,722. However, the accretion luminosity of FU Ori has remained high (at 10-100 times higher accretion rate than HBC\,722) for 75 years, whereas HBC\,722 appears to be peaking after only a few years.  V1057 Cyg (at $\sim$ 40 years post-outburst) is surrounded by considerably more nebulosity than FU Ori or HBC\,722, but still shows deep blueshifted absorption features. 

It has been argued for V1057 Cyg \citep{petrov98} and FU Orionis \citep{petrov08} that the lack of difference in the width of the optical lines indicates that the boxy/double-peaked profiles  originate from the central star only.  In contrast, we attribute only the narrow central absorption to the central object, and the ``wings'' of the features to the disk.  The detection of a separate stellar absorption in HBC\,722 is possible because HBC\,722 is a much less luminous system (5-12 \Lsun) than FU Ori or V1057 Cyg (100 - 250 \Lsun) with a correspondingly lower inferred accretion rate.  If these features are disk-driven in the more luminous objects, a similar narrow absorption may appear as they fade.

Because we observed HBC\,722 over multiple epochs, this dataset is the first opportunity to see the line profiles change as the burst evolves.  The double-peaked feature in most lines was weak before August 2012, when the wind was still strong and dominated the absorption profiles.  The broad blueshifted (wind) component almost disappeared in August of 2012; at the same time, the double-peaked profiles emerged and stabilized.  

\begin{figure}[!p]
\begin{center}
\epsscale{1.0}
\plotone{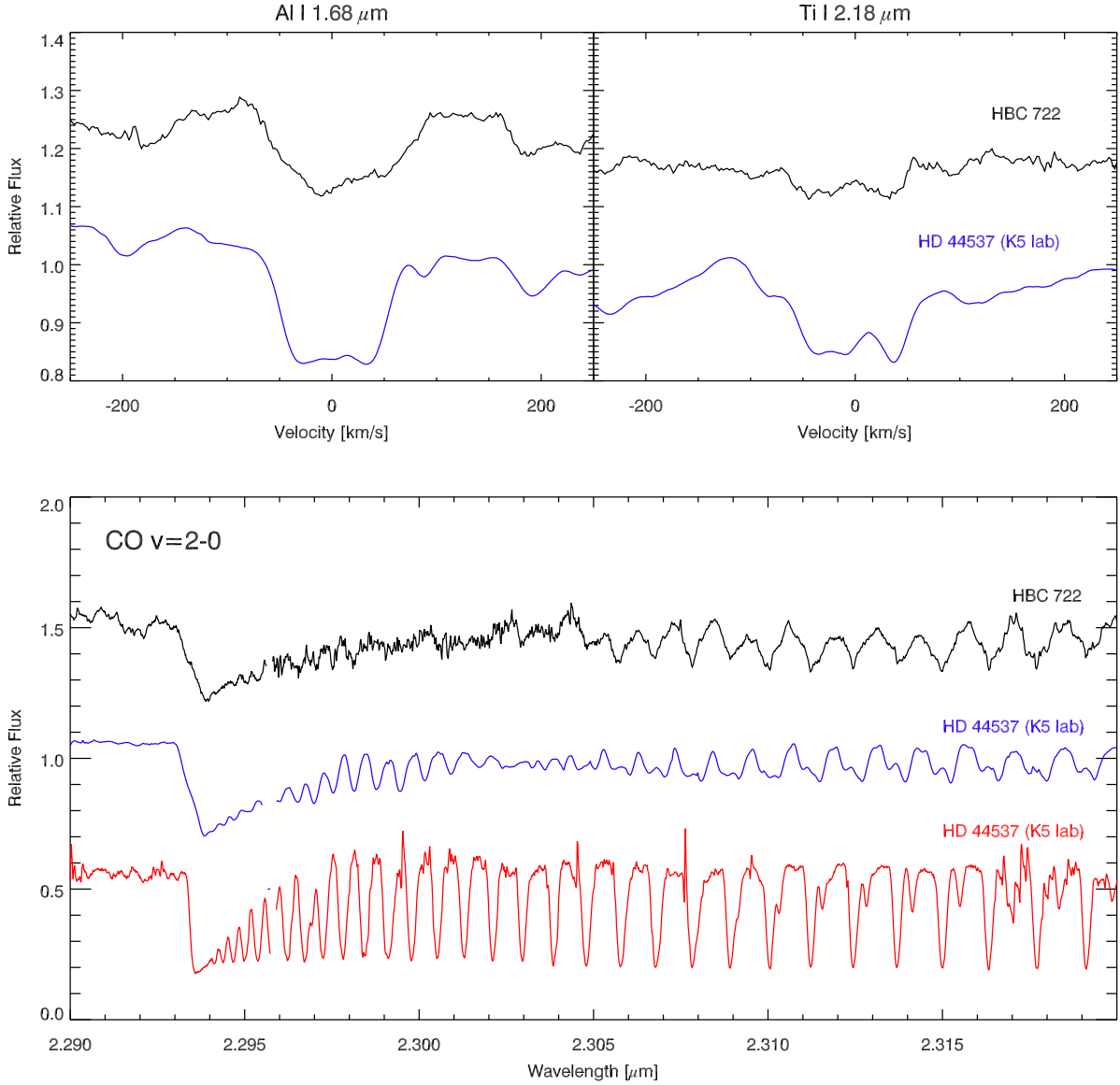}
\caption{IGRINS spectra (presented in black) of neutral metal lines and CO overtone transitions that show a double-peaked profile. The blue spectrum is from a template K5-type (HD 44537) stellar spectrum convolved with a rotational velocity of 50 \kms.  The original stellar spectrum of HD 44537 is shown in red. 
} 
\label{igrins}
\end{center}
\end{figure}

We also observed a single epoch of HBC\,722 with the high-resolution IGRINS spectrograph in 2014 (Figure \ref{igrins}). We compare spectra of HBC\,722 (black) with a K5-type stellar spectrum (HD 44537) convolved with a rotational profile of $vsini=50$ \kms (blue). The narrow stellar component is clearly seen at the line center of the CO overtone transitions.

\begin{figure}[!p]
\begin{center}
\epsscale{1.0}
\plotone{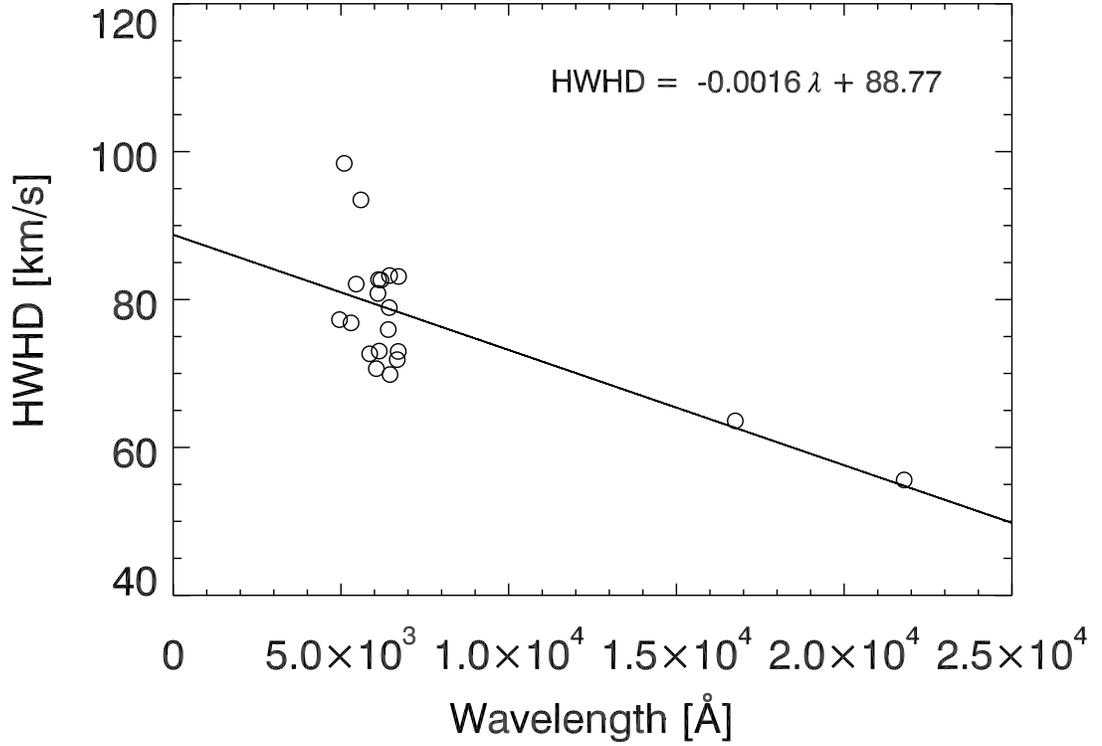}
\caption{The HWHD of spectra with a double-peaked absorption feature (see Table 2) as a function of wavelength. We observe a decrease in HWHD with wavelength, consistent with Keplerian disk rotation.
The Pearson correlation coefficient between HWHD and wavelength is -0.69 (p-value=0.0008), and the Kendall's tau correlation coefficient is -0.32 (p-value=0.05).   
The relation presented at the upper right is the result of linear regression (the solid line).
} 
\label{hwhd}
\end{center}
\end{figure}

Figure \ref{hwhd} presents the HWHD of absorption spectra as a function of wavelength.  In order to increase the signal-to-noise ratio, we averaged the last three epochs of optical spectra (30 August 2012, 30 May 2013, and 08 August 2013). The IGRINS spectra were observed only once.  The HWHD decreases with wavelength, particularly in the IR. This is consistent with Keplerian disk rotation; longer wavelengths trace the cooler outer part of the disk where the rotation velocity is
smaller.
The Pearson correlation coefficient between HWHD and wavelength is -0.69 (p-value=0.0008), which is within 95 \% confidence interval.

\section{Discussion}

Stars form by the accretion of material from disks to protostars, and this accretion process often triggers a disk wind.  We observe the wind as a blueshifted absorption feature or as a P Cygni profile in optical lines \citep{hartmann96}.  Our high resolution optical spectra clearly show a wind feature in HBC\,722, which appeared as soon as the object was observed after its optical burst (2010). The wind velocity peaked at about 400 \kms, and sustained for $\sim$2 years.  It disappeared almost completely during summer 2012, although the brightness continued to rise into the first half of 2013 and stayed at a constant level after then. At the highest observed velocity, the wind could have traveled up to $\sim$170 AU during those 2 years.

The second set of features are the blueshifted and redshifted boxy/double-peaked absorption profiles, which we attribute to a self-luminous disk rotating at Keplerian speeds.  This interpretation is supported by the decrease in HWHD of the double-peaked profile from optical to near-IR (Figure \ref{hwhd}), and by the presence of a central narrow absorption feature that can be attributed to the central object.  The optical lines have both greater rotational velocities ($\sim$ 70 \kms\ compared with 50 \kms\ in the IR) and are matched by an earlier spectral type template (G5) than the IR lines (K5), as presented with blue spectra in Figure~\ref{doublepeak} and Figure~\ref{igrins}.

If HBC\,722 has a Keplerian disk and its central protostellar mass is 1 \Msun, the double-peaked optical spectral trace the disk rotating at 39 \Rsun\ while the near-IR spectra traces the 
disk rotating at 76 \Rsun.  This indicates that the temperature of the disk decreases from $\sim$5,200 K at the radius of 39 \Rsun\ to $\sim$3000 K at the radius of 76 \Rsun. 
If the power-law temperature distribution is adopted, this decrease of temperature corresponds to 
the power-law index of 0.8, which is very similar to $\frac{3}{4}$ for the optically thick disk with a steady accretion \citep{hartmann96}.
Both temperatures are greatly enhanced over a quiescent T Tauri-type disk at corresponding radii, indicating that the disk has remained hot despite the wind turnoff in 2012.  

The bolometric luminosity of the system reached and stayed at its second maximum for at least 2 years beyond the wind turnoff time, which was itself 2 years after the initial rise (which lasted only 3 months).  This suggests the initial optical burst was driven by a phenomenon with a few month timescale (perhaps a free fall time or a few orbital times at the disk inner edge), while the driving source of the wind was a longer timescale event, perhaps an accretion stream from further out in the disk.

One interesting phenomenon is that the disk feature started to appear when the wind feature disappeared.  This phenomenon might hint at the a rebuilding timescale for the inner disk. 
After a burst accretion event, the inner disk can be rebuilt by the material infalling from the outer disk or an extended envelope, depending on the evolutionary stage of the outbursting object. 
However, wind pressure could prevent the material from refilling immediately, maintaining the inner boundary of the disk at a relatively large radius.  Once the wind turns off, the disk material can move inward once again to rebuild the inner hot disk, which is then visible at optical wavelengths once more.  High spatial resolution observations of the inner disk of HBC\,722 from telescopes such as ALMA might provide strong constraints on disk accretion and rebuilding processes. 

\section{Conclusions}

We have observed the burst of HBC\,722 in unprecedented detail with high resolution optical and near-IR spectroscopy over multiple epochs.  The optical line profile of \ha\ reveals a blueshifted 400 \kms\ wind immediately after burst.  As the system luminosity (and therefore accretion rate) of the system increased from 2010 - 2012, the wind remained strong.  In 2012, the wind diminished and double-peaked rotating disk profiles (70 \kms, spectral type G5) appeared in the neutral metal lines.  In 2013, a narrow central absorption, likely the central star, appeared in the neutral metal lines.  In 2014, the IR line profiles (including CO overtone features) were observed to be double-peaked but with a lower rotational velocity (50 \kms) and later spectral type (K5), suggesting that they arise farther out in the disk than the optical lines.  Combined with high resolution interferometry and additional optical/IR spectral coverage as the burst evolves, this dataset will reveal new constraints on the accretion and outflow process in the inner disks of low mass protostars.

\acknowledgements
This work was supported by a grant from the Kyung Hee University in 20130384.
This work was also supported by the Basic Science Research Program (grant No. NRF-2012R1A1A2044689) and the BK21 plus program through the National Research Foundation (NRF) funded by the Ministry of Education of Korea. J.-E. L. is very grateful to the department of Astronomy, University of Texas
at Austin for the hospitality provided to her from August 2013 to July 2014.  JDG thanks Aditi Raye Allen for early data reduction and analysis of the HET data.
The Hobby-Eberly Telescope (HET) is a joint project of the University of
Texas at Austin, the Pennsylvania State University, Stanford University,
Ludwig-Maximilians-Universit\"{a}t M\"{u}nchen, and
Georg-August-Universit\"{a}t G\"{o}ttingen. The HET is named in honor
of its principal benefactors, William P. Hobby and Robert E. Eberly.
This work used the Immersion Grating Infrared Spectrograph (IGRINS) that was developed under a collaboration between the University of Texas at Austin and the Korea Astronomy and Space Science Institute (KASI) with the financial support of the US National Science Foundation under grant AST-1229522, of the University of Texas at Austin, and of the Korean GMT Project of KASI.

\clearpage

\clearpage


\begin{deluxetable}{lccllc}
\footnotesize
\tabletypesize{\scriptsize}
\tablecaption{Observing log \label{tbl_obs}}
\tablewidth{0pt}
\tablehead{
  \colhead{Telescope} & \colhead{Instrument} & \colhead{resolving power (R)} & \colhead{target} & \colhead{Date} & \colhead{Exposure time}  \\
  \colhead{} & \colhead{(wavelength coverage)} & \colhead{} &  \colhead{} &\colhead{} & \colhead{[sec]} 
  } 
\startdata
HET & HRS &  30,000 &HBC 722 & 2010 Dec 07 & 3600 \\
 & (4790 --6810 \AA) &  & HBC 722 & 2011 May 31 & 3600 \\
&	&  &HBC 722 & 2012 Aug 08 & 3600 \\
&	&  &HBC 722 & 2012 Aug 30 & 3600 \\
&	&  &HBC 722 & 2013 Aug 08 & 3600 \\
 \cline{1-6}
BOAO & BOES & 30,000&HBC 722 & 2010 Nov 26	& 3600 \\
& (3600 --10500 \AA) &	&  HBC 722 & 2010 Dec 11	& 3600 \\
& 	&  &HBC 722 & 2010 Dec 23	& 3600 \\
& 	&  &HBC 722 & 2010 Dec 29	& 3600 \\
& 	&  &HBC 722 & 2011 Jan 20	& 3600 \\
& 	&  &HBC 722 & 2011 Apr 20	& 3600 \\
& 	&  &HBC 722 & 2011 May 18	& 3600 \\
& 	&  &HBC 722 & 2011 Jun 16	& 3600 \\
& 	&  &HBC 722 & 2011 Sep 06	& 3600 \\
& 	&  &HBC 722 & 2011 Sep 25	& 3600 \\
& 	&  &HBC 722 & 2012 May 21  	& 3600 \\
& 	&  &HBC 722 & 2012 Sep 10	& 3600 \\ 
& 	&  &HBC 722 & 2013 Apr 21	& 3600 \\
& 	&  &FU Ori & 2011 Jan 26	& 3600 \\
& 	&  &V1057 Cyg & 2012 Sep 11	& 3600 \\
&	& &HD 52497 & 2011 Sep 25	& 300 \\
&	& &HD 188650 & 2011 Sep 22	& 600 \\
 \cline{1-6}
HJST & IGRINS & 40,000 & HBC 722 & 2014 Nov 20 & 2400 \\
&(1.47--1.81 $\mu$m, 1.95--2.48 $\mu$m) & &HIP 103108 & 2014 Nov 20 & 1200 \\ 
&      & &HD 44537 & 2014 Nov 17 & 8\\
\enddata
\end{deluxetable}

\clearpage


\begin{deluxetable}{lccc}
\footnotesize
\tabletypesize{\scriptsize}
\tablecaption{List of Double-Peaked Lines\label{tbl_hwhm}}
\tablewidth{0pt}
\tablehead{\colhead{$ \lambda~[\AA]$}& \colhead{Ion}& \colhead{HWHD [km s$^{-1}$]}}
  \startdata
 4957.60 & Fe I & 77.28 & \\
 5098.70 & Fe I & 98.41 & \\
 5302.30 & Fe I & 76.84 & \\
 5455.61 & Fe I & 82.09 & \\
 5594.47 & Ca I & 93.47 & \\%
 5857.45 & Ca I & 72.66 & \\%
 6056.01 & Fe I & 70.65 & \\%
 6102.72 & Ca I & 80.83 & \\%
 6122.22 & Ca I & 82.70 & \\%
 6141.73 & Fe I & 73.03 & \\%
 6191.56 & Fe I & 82.64 & \\%
 6411.65 & Fe I & 75.94 & \\ %
 6439.08 & Ca I & 78.91 & \\%
 6449.81 & Ca I & 83.24 & \\%
 6462.57 & Ca I & 69.86 & \\%
 6677.99 & Fe I & 71.87 & \\%
 6707.89 & Li I & 72.98 & \\%
 6717.68 & Ca I & 83.12 & \\%
 16755 & Al I & 63.59 & \\
 21789 & Ti I & 55.60 & \\
\enddata
\end{deluxetable}

\clearpage 

\end{document}